\documentstyle[aps,prl,epsf,epsfig,floats,twocolumn]{revtex}
\begin{document}
\draft
\twocolumn[\hsize\textwidth\columnwidth\hsize\csname @twocolumnfalse\endcsname
\title{Onset of Glassy Dynamics in a Two-Dimensional Electron System in 
Silicon}
\author{Sne\v{z}ana Bogdanovich and Dragana Popovi\'{c}}
\address{National High Magnetic Field Laboratory, 
Florida State University, Tallahassee, FL 32310 }
\date{\today}
\maketitle

\begin{abstract}
The time-dependent fluctuations of conductivity $\sigma$ have been 
studied in a two-dimensional electron system in low-mobility, small-size Si 
inversion layers. The noise power spectrum is $\sim 1/f^{\alpha}$ with 
$\alpha$ exhibiting a sharp jump at a certain electron
density $n_s=n_g$.  An enormous increase in the relative variance of $\sigma$ 
is observed as $n_s$ is reduced below $n_g$, reflecting a dramatic slowing down
of the electron dynamics.  This is attributed to the freezing of the electron 
glass.  The data strongly suggest that glassy dynamics persists in the 
metallic phase.
\end{abstract}

% 71.30.+h Metal-insulator transitions...
% 71.27.+a Strongly correlated electron systems
% 73.40.Qv Electronic transport in metal-insulator-semiconductor structures
]
	
The possibility of a metal-insulator transition (MIT) in two dimensions (2D) 
has been a subject of intensive research in recent 
years~\cite{Abrahams,Feng} but the physics behind this phenomenon is still not
understood. It is well established that the MIT occurs in the regime where 
both Coulomb 
(electron-electron) interactions and disorder are strong.  Theoretically, it 
is well known~\cite{hopping} that, in the strongly localized limit, the 
competition between electron-electron interactions and disorder leads to 
glassy dynamics (electron or Coulomb glass).  Some glassy properties, such as 
slow relaxation phenomena, have been indeed observed in various insulating 
thin films~\cite{films2,films3,films4}.  Furthermore, recent 
work~\cite{newgang} has suggested that the critical behavior near the 2D MIT 
may be dominated by the physics of the insulator, leading to the proposals 
that the 2D MIT can be described alternatively as the melting of the Wigner 
glass~\cite{Chakra}, or the melting of the electron glass~\cite{Pastor}.  It 
is clear that understanding the nature of the insulator represents a major 
open issue in this field.  Here we report the first detailed study of glassy 
behavior in a 2D system in semiconductor heterostructures.  The glass 
transition is manifested by a very abrupt onset of a specific type of 
random-looking slow dynamics, together with other signs of cooperativity.  Our
results strongly suggest that the glass transition occurs in the metallic 
phase as a precursor to the MIT, in agreement with recent 
theory~\cite{Dobrosavljevic}.

While glassy systems exhibit a variety of phenomena~\cite{glasses}, studies of
metallic spin glasses have demonstrated~\cite{Weissman} that mesoscopic, {\it
i.~e.} transport noise measurements are required in order to provide 
definitive 
information on the details of glassy ordering and dynamics.  Fluctuations of
conductivity $\sigma$ as a function of chemical potential (or gate voltage 
$V_g$, which controls the carrier density $n_s$) have been investigated 
extensively in the insulating regime~\cite{Fowler_fluct} and near the 
MIT~\cite{DP_meso} in a 2D electron system in mesoscopic Si 
metal-oxide-semiconductor field-effect transistors (MOSFETs).  The latter 
study indicated that Coulomb interactions dominate the physics near the MIT.  
In order to get reasonably reproducible fluctuations as a function of $V_g$, 
it was necessary to make very slow sweeps of many hours over a very narrow 
range of $V_g$.  Thus, all measurements represented a time average.  As a
matter of fact, it had been already known~\cite{Fowler_time} that, at fixed 
$V_g$ (or $n_s$), $\sigma$ fluctuates as a function of time.  Both high- 
and low-frequency fluctuations were evident.  It was speculated that the time 
dependence of $\sigma$ was due to correlated transitions of electrons between 
different configurations~\cite{Fowler_time} or, in other words, between 
different metastable states in the glassy phase, but there has been no detailed
study of these effects up to now.  Here we present the first systematic study 
of transport and noise in a strongly disordered, mesoscopic 2D system over a 
wide range of $n_s$ and $T$.

Most of the measurements were carried out on a 
1~$\mu$m long, 90~$\mu$m wide rectangular n-channel Si MOSFET with the peak 
mobility of only 0.06~m$^2$/Vs at 4.2~K (with the applied back-gate bias of 
$-2$~V). The samples were fabricated using standard $0.25~\mu$m Si 
technology~\cite{Taur} with poly-Si gates, self-aligned ion-implanted 
contacts, substrate doping $N_a\sim 2\times10^{17}$cm$^{-3}$, oxide charge 
$N_{ox}=1.5\times10^{11}$cm$^{-2}$, and oxide thickness $d_{ox}=50~nm$. 
The fluctuations of current $I$ ({\it i.~e.}
$\sigma$) were measured as a function of time in a two-probe configuration 
using an ITHACO 1211 current preamplifier and a PAR124A lock-in amplifier at
$\sim 13$~Hz.  The excitation voltage $V_{exc}$ was kept constant and low 
enough (typically, a few $\mu$V) to ensure that the conduction was Ohmic.  A
precision DC voltage standard (EDC MV116J) was used to apply $V_g$.  The 
current fluctuations as low as $10^{-13}$~A were measured at $0.13\leq T\leq 
0.80$~K in a dilution refrigerator with heavily filtered wiring.  Relatively
small fluctuations of $T$, $V_g$, and $V_{exc}$ were ruled out as possible 
sources of the measured noise, since no correlation was found between them and
the current fluctuations.  In addition, a Hall bar sample from the same wafer 
was measured at $T=0.25$~K in both two- and four-probe configurations, and it 
was determined that the contact resistances and the contact noise were 
negligible.

\begin{figure} 
\vspace*{-0.3in}
\epsfxsize=3.0in \epsfbox{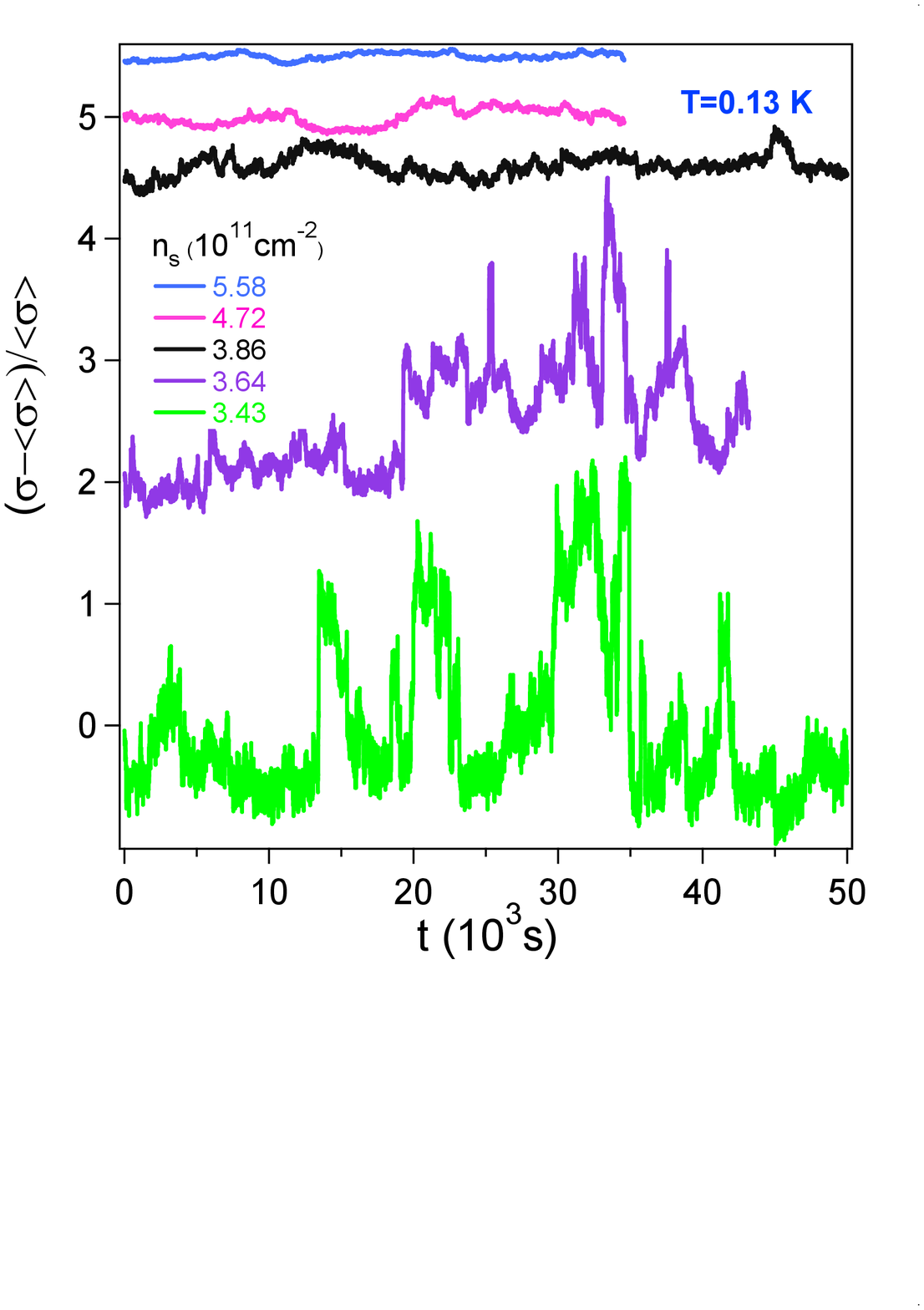}
\vspace*{-1.05in}
\caption{Relative fluctuations of $\sigma$ {\it vs.} time for different $n_s$ 
at $T=0.13$~K.  Different traces have been shifted for clarity, starting with 
the lowest $n_s$ at bottom and the highest at top.
\vspace*{-0.2in}
\label{data}}
\end{figure}
The relative fluctuations $(\sigma-\langle\sigma\rangle)/\langle\sigma\rangle$
(averaging over time
intervals of several hours) are shown in Fig.~\ref{data} for a few selected 
$n_s$ at $T=0.13$~K.  It is quite striking that, for the lowest $n_s$, the 
fluctuation amplitude is of the order of 100~\%.  In addition to rapid, 
high-frequency fluctuations, both abrupt jumps and slow changes over periods 
of several hours are also evident.  The amplitude of the fluctuations 
decreases with increasing either $n_s$ or $T$, as discussed in more detail 
below.
\begin{figure}
\vspace*{-0.5in}
\epsfxsize=3.0in \epsfbox{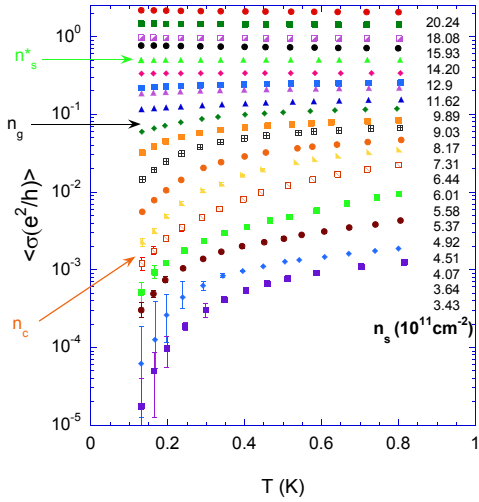}
\vspace*{-0.5in}
\caption{$\langle\sigma\rangle$ {\it vs.} $T$ for different $n_s$.  The data 
for many other $n_s$ have been omitted for clarity.  The error bars show the 
size of the fluctuations.  $n_{s}^{\ast}$, $n_g$, and $n_c$ are marked by 
arrows.  They were determined as explained in the main text.
\vspace*{-0.2in}
\label{saverage}}
\end{figure}

Figure~\ref{saverage} shows the time-averaged conductivity 
$\langle\sigma\rangle$ as a function of $T$ for different $n_s$.  
The behavior of $\langle\sigma (n_s,T)\rangle$ in our samples is 
found to be somewhat similar to that of high-mobility Si MOSFETs.  At the 
highest $n_s$, for example, our devices exhibit a metalliclike behavior with 
$d\langle\sigma\rangle/dT<0$.  The change of $\langle\sigma\rangle$ in a given
$T$ range, however, is 
small (only 6\% for the highest $n_s=20.2\times10^{11}$cm$^{-2}$) as observed 
in other Si MOSFETs with a large amount of disorder~\cite{Pudalov_drop,WL}.  
$d\langle\sigma\rangle/dT$ changes sign when 
$\langle\sigma(n_{s}^{\ast})\rangle=0.5~e^{2}/h$ similar to other 2D 
samples~\cite{Abrahams}.  Even though the
corresponding density $n_{s}^{\ast}=12.9\times10^{11}$cm$^{-2}$ is much higher
due to a large amount of disorder in our devices, the effective Coulomb
interaction is still comparable to that in other 2D systems ($r_s\sim 4$, 
$r_s$--ratio of Coulomb energy to Fermi energy).

\begin{figure}[t]
\epsfxsize=3.2in \epsfbox{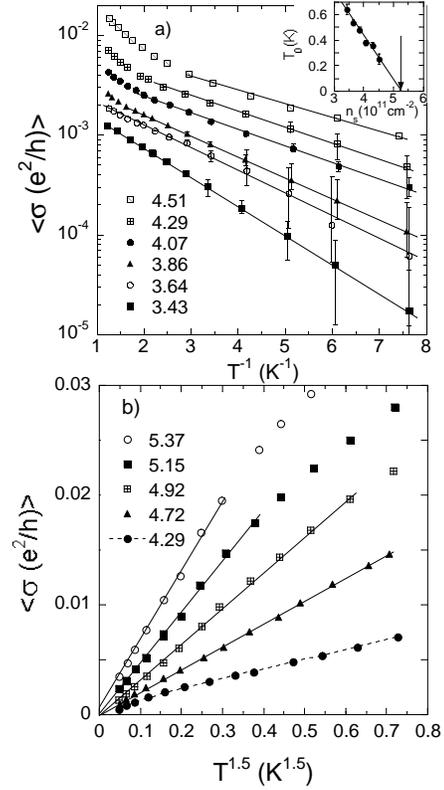}
\vspace*{0.1in}
\caption{a) $\langle\sigma\rangle$ {\it vs.} $T^{-1}$ for several 
$n_s (10^{11}$cm$^{-2})$ in the insulating regime. The error bars 
show the size of the fluctuations, and the lines are fits to 
$\langle\sigma\rangle\propto\exp(-T_0/T)$.  Inset: $T_0$ {\it vs.} $n_s$ with 
a linear fit, and an arrow showing $n_c$. b) $\langle\sigma\rangle$ 
{\it vs.} $T^{1.5}$ for a few $n_s (10^{11}$cm$^{-2})$ near $n_c$.  
The solid lines are fits; the dashed line is a guide to the eye, clearly 
showing insulating behavior ($\langle\sigma(T\rightarrow 0)\rangle=0$).
\label{act}}
\end{figure}

The density $n_{s}^{\ast}$, where $d\langle\sigma\rangle/dT=0$, has been
usually~\cite{Abrahams} identified with the critical density for the MIT.  In
high-mobility Si MOSFETs, the critical density has been also 
determined~\cite{activated} as the density $n_c$ where activation energy 
associated with the insulating exponential behavior of $\langle\sigma 
(T)\rangle$ vanishes.  It was established that $n_{s}^{\ast}\approx n_c$, 
although a small but systematic difference of a few percent has been 
reported~\cite{Pudalov_drop,Altshuler_weakloc} such that $n_{s}^{\ast}>n_c$.
For the lowest $n_s$ in our experiment, the data are best 
described by the simply activated form $\langle\sigma\rangle\propto\exp 
(-T_{0}/T)$ 
[Fig.~\ref{act}(a)], consistent with other studies close enough to the 
MIT~\cite{activated}.  The data could not be fitted satisfactorily to any
variable-range hopping (VRH) law regardless of the prefactor~\cite{vrh_foot}.
$T_0$ decreases linearly with increasing $n_s$ (Fig.~\ref{act}(a) inset), and 
vanishes at $n_c\approx5.2\times10^{11}$cm$^{-2}$.  
Close to $n_c$, the data are best described by the metallic 
power-law behavior $\langle\sigma(n_s,T)\rangle=a(n_s)+b(n_s)T^{x}$ with 
$x\approx1.5$
[Fig.~\ref{act}(b)].  The fitting parameter $a(n_s)$ is relatively small
and, in fact, vanishes for $n_s (10^{11}$cm$^{-2})=4.72$ and 4.92.  Such a 
simple power-law $T$-dependence of $\sigma$, given by $\langle\sigma 
(n_c,T)\rangle\propto T^x$,
is consistent with the one expected in the quantum critical region (QCR) of 
the MIT based on general arguments~\cite{Belitz}, and with the behavior 
observed in 3D systems~\cite{Belitz} and other Si MOSFETs~\cite{Feng} within 
the QCR.  Therefore, based on the analysis of 
$\langle\sigma (n_s,T)\rangle$ in both insulating regime and QCR,
we conclude that
the critical density $n_c=(5.0\pm 0.3)\times 10^{11}$cm$^{-2}$ ($r_s\sim 7$), 
which is more than a factor of two smaller than $n_{s}^{\ast}$. Such a large 
difference between $n_c$ and $n_{s}^{\ast}$ is attributed to a much higher 
amount of disorder in our samples than in high-mobility Si 
MOSFETs~\cite{Pudalov_drop,Altshuler_weakloc,activated}.

\begin{figure}
\vspace*{-0.5in}
\epsfxsize=2.9in \epsfbox{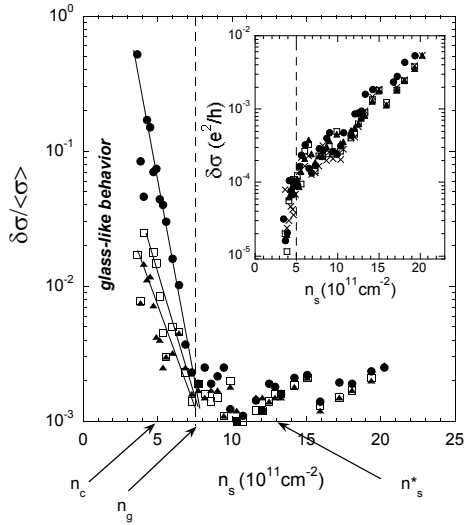} 
\vspace*{-0.35in}
\caption{$\delta\sigma/\langle\sigma\rangle$ (main) and $\delta\sigma$ (inset)
{\it vs.} $n_s$ at different $T$ ($\bullet$: 0.130 K, $\times$: 0.196 K, 
$\Box$: 0.455 K, triangle: 0.805 K).  Main: $n_{s}^{\ast}$, $n_g$, and $n_c$ 
are marked by arrows.  The vertical dashed line shows the region of densities 
$n_s<n_g$ where various glassy properties have been observed.  Inset: The 
vertical dashed line shows the location of the critical density $n_c$.
\label{rms}}
\end{figure}

The fluctuations of $\sigma$ with time have been studied first by analyzing 
$\delta\sigma=\langle(\sigma-\langle\sigma\rangle)^2\rangle ^{1/2}$.  
Fig.~\ref{rms} inset shows that,
while $\delta\sigma$ does not seem to depend on $T$, it increases with $n_s$ 
by three orders of magnitude.  The most striking feature of the data, however,
is the sudden and dramatic change in the rate of its $n_s$ dependence (see
``kink'' in Fig.~\ref{rms} inset), which occurs near $n_c$.  Although we are 
not aware of any theoretical work relevant to this problem, we note that the
observed $\delta\sigma (n_s)$ is plausible: once electrons enter the localized
({\it i.~e.} insulating) phase by reducing $n_s$ below $n_c$, their ability 
to change configurations will be severely impaired, resulting in a much more 
rapid drop of the fluctuation amplitude $\delta\sigma$ with decreasing $n_s$.

The main part of Fig.~\ref{rms} shows that, for high $n_s$, the relative 
amplitude of fluctuations $\delta\sigma/\langle\sigma\rangle$ is independent 
of $n_s$ and $T$.  However, below a certain density $n_g = 
(7.5\pm 0.3)\times10^{11}$cm$^{-2}$, which does not seem to depend on $T$, an 
enormous increase of $\delta\sigma/\langle\sigma\rangle$ is observed with 
decreasing either $n_s$ or $T$.  It is interesting that 
$\delta\sigma/\langle\sigma\rangle$ does not
exhibit any special features near $n_c$ or $n_{s}^{\ast}$.  The onset of strong
noise at $n_g>n_c$ is, in fact, consistent with the observation~\cite{Cohen}
that, in some materials, a considerable increase in noise occurs in the
metallic phase as a precursor to the MIT.  We show below that 
here $n_g$ represents the density below which the 2D electron system freezes 
into an electron glass. 

\begin{figure*}
\vspace*{-0.65in}
\epsfig{file=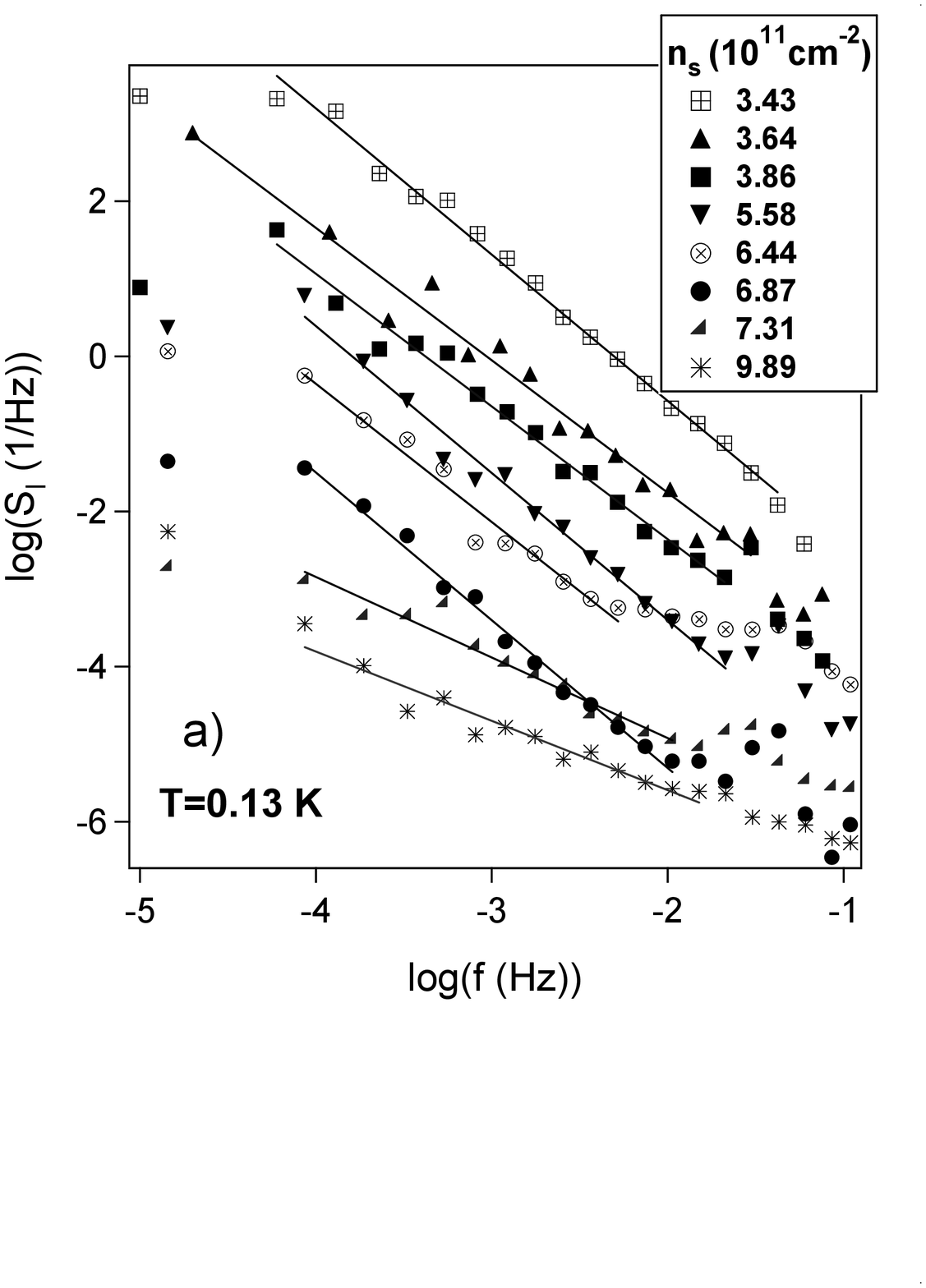,width=2.22in,clip=}\hspace*{-0.2in}
\epsfig{file=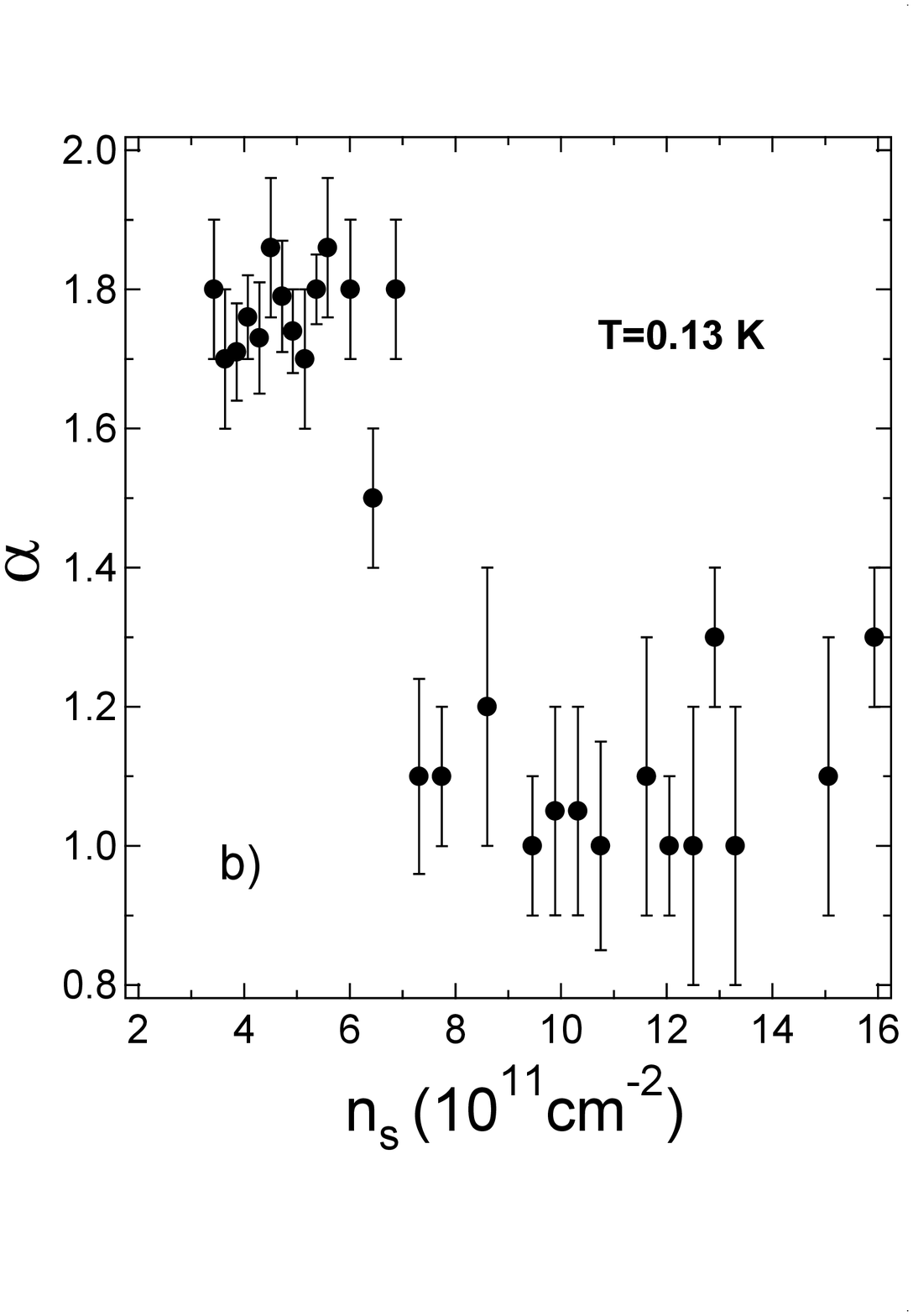,width=2.25in,bbllx=11,bblly=-35,bburx=601,bbury=776,clip=}\hspace*{0.0in}
\epsfig{file=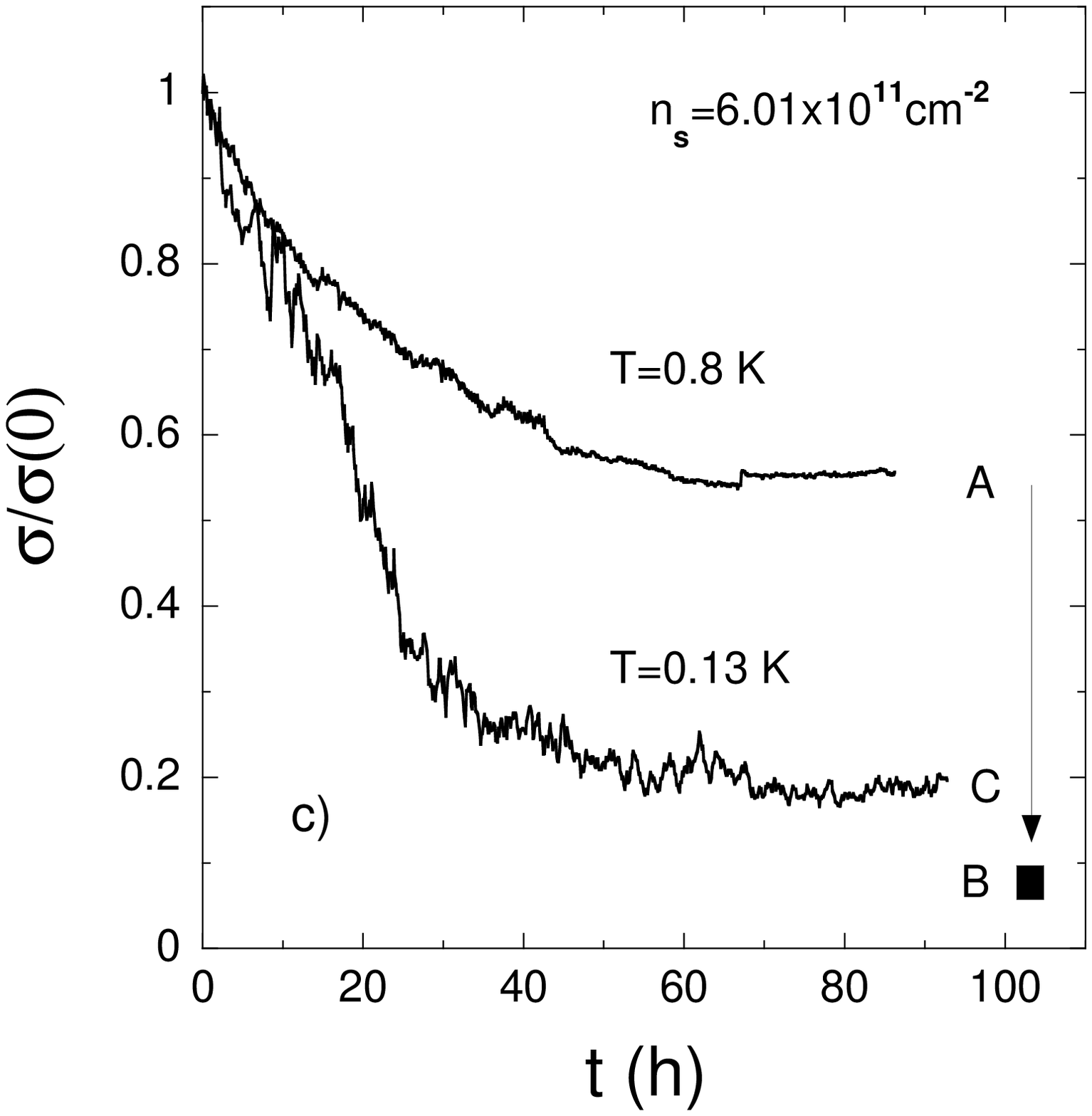,width=2.48in,bbllx=11,bblly=-49,bburx=601,bbury=776,clip=}
\vspace*{-0.6in}
\caption{(a) The averaged noise power spectra $S_I\propto 1/f^{\alpha}$ {\it 
vs.} $f$ for several $n_s$. Solid lines are linear least-squares fits 
with the slopes equal to $\alpha$.  (b) $\alpha$ {\it vs.} $n_s$.  (c)  
Relaxation of $\sigma$ following a slow change in $n_s (10^{11}$cm$^{-2})$ 
from 15.93 to 6.01, carried out at $T=0.8$ and $T=0.13$~K over a period of 
$\approx 4.5$ hours each.  After $\sigma$ reached a stationary value (A) at 
0.8~K, the sample was cooled down to 0.13~K.  The resulting $\sigma$ (B)
differs from $\sigma$ (C) obtained using a different cooling procedure by a
factor of two, clearly demonstrating history dependent, {\it i.~e.} non-ergodic
behavior.  Such behavior is not observed for $n_s>n_g$.
\label{sf}}
\end{figure*}

The normalized power spectra 
$S_{I}(f)=S(I,f)/I^{2}$ ($f$--frequency) of 
$(\sigma-\langle\sigma\rangle)/\langle\sigma\rangle$ were also studied for all
$n_s$ and $T$.  
Most of the spectra were obtained in the $f=(10^{-4}-10^{-1})$~Hz bandwidth, 
where they follow the well-known empirical law
$S_I=\beta/f^{\alpha}$~\cite{Hooge,Weissman1/f}
($\beta$ is inversely proportional to the number of fluctuators and, usually,
$\alpha\sim 1$).
The background noise was measured by setting $I=0$ for all $n_s$ and $T$.  It 
was always white and usually several orders of magnitude smaller than the 
sample noise.  Nevertheless, a subtraction of the background spectra was 
always performed, and the power spectra of the device noise were averaged over
frequency bands ($\lesssim$ an octave).  Some of the resulting $S_I$
are presented in Fig.~\ref{sf}(a).
At the highest $n_s$ (not shown), $S_I(f)$ does not depend on $n_s$.  However,
it is obvious that, by reducing $n_s$ below $n_g$, $S_I$ increases enormously,
by up to six orders of magnitude at low $f$.  This striking increase of the 
slow dynamic contribution to the conductivity is consistent with the behavior 
of $\delta\sigma/\langle\sigma\rangle$ (Fig.~\ref{rms}).  In fact, since 
relative variance $(\delta\sigma)^2/\langle\sigma\rangle^2=\int S_I(f)df$, it
is clear that the observed giant increase of 
$\delta\sigma/\langle\sigma\rangle$ for $n_s<n_g$ (Fig.~\ref{rms}) reflects a 
{\em sudden and dramatic slowing down of the electron dynamics}.  This is 
attributed to the freezing of the electron glass.
We also find that, for $n_s<n_g$, $S_I(f)$ {\em increases} exponentially with 
decreasing $T$. The observed $T$-dependence of noise (obvious from 
Fig.~\ref{rms}) is consistent with early studies on Si MOSFETs~\cite{Adkins},
and it shows that the noise in our system cannot be explained by the models of
thermally activated charge trapping~\cite{Weissman1/f,Dutta,Buhrman}, noise 
generated by fluctuations
of $T$~\cite{Voss}, noise in the hopping regime~\cite{hopping_noise}, and in
the vicinity of the Anderson transition~\cite{Cohen}.  On the other hand, 
similar increase of noise at low $T$ has been observed in mesoscopic spin
glasses~\cite{Israeloff,Jaroszy}, in wires in the quantum Hall regime for 
tunneling through localized states~\cite{Wrobel}, and in Si quantum dots in the
Coulomb blockade regime~\cite{Molenkamp}.

Another remarkable result, shown in Fig.~\ref{sf}(b), is a sharp jump of the
exponent $\alpha$ at $n_s\approx n_g$.  While $\alpha\approx 1$ for $n_s>n_g$,
$\alpha\approx 1.8$ below $n_g$, reflecting a sudden shift of the spectral 
weight towards lower frequencies.  Similar large values of $\alpha$ have been 
observed in spin glasses above the MIT~\cite{Jaroszy}, and in submicron 
wires~\cite{Wrobel}.  In general, such noise with spectra closer to $1/f^2$ 
than to $1/f$ is typical of a system far from equilibrium, in which a 
step does not lead to a probable return step.  We also have the analysis of 
higher order statistics (non-Gaussianity or second 
spectra~\cite{Weissman,Weissman1/f}) of the noise, showing an abrupt change to 
the sort of statistics characteristic of complicated multi-state systems just 
at the density $n_g$ at which $\alpha$ jumps.  This will be described in 
detail elsewhere.

We have demonstrated that the transition to a glassy phase is characterized by
a sudden, enormous increase in the low-frequency noise in $\sigma$, a sudden 
shift of the spectral weight towards lower $f$, and a dramatic 
increase of noise with decreasing $T$.  Similar behavior in spin glasses was 
attributed to spin glass freezing~\cite{Israeloff,Jaroszy}.  In addition, for 
$n_s<n_g$, we have observed long relaxation times following a large change in 
$V_g$, and history dependent behavior characteristic of a glassy phase 
[Fig.~\ref{sf}(c)].  
%These effects will be described in detail elsewhere.  Here 
We note that in order to obtain reproducible values of 
$\langle\sigma (n_s,T)\rangle$ shown in Fig.~\ref{saverage}, it was necessary
to vary $n_s$ in small steps of $4.3\times 10^{10}$cm$^{-2}$ at the highest 
$T$ (0.8~K).  

In summary, we present the first evidence of electron glass freezing at a
well-defined density $n_g$ in a 2D electron system in silicon, in agreement 
with theoretical predictions~\cite{Chakra,Pastor}.  Glassy freezing occurs in 
the regime of very low $\langle\sigma\rangle$, apparently as a precursor to 
the MIT.  The existence of such an intermediate ($n_c<n_s<n_g$) glass phase is 
consistent with theoretical predictions~\cite{Dobrosavljevic}. 

The authors are grateful to the Silicon Facility at IBM, Yorktown Heights for 
sample fabrication, and to V. Dobrosavljevi\'{c} and J. Jaroszy\'{n}ski for 
useful discussions.  This work was supported by NSF Grant DMR-0071668 and by 
an NHMFL In-House Research Program grant.


\begin{references}
\bibitem{WL} It is interesting that weak localization (WL) has not been seen 
in $\sigma (T)$ even though our lowest $T\sim 9\times 10^{-4}T_F$ (Fermi 
temperature $T_F$[K]$=7.31\times n_s$[$10^{11}$cm$^{-2}$])
is almost an order of magnitude 
lower than $T$ below which WL effects are expected to set 
in~\cite{Pudalov_weakloc,Altshuler_weakloc}.
\bibitem{vrh_foot} We expect that VRH would become important at even lower 
$n_s$, which could not be reached experimentally because of the low 
signal-to-noise ratio.
\end{references}
\end{document}